%% file: Kaisin_n_en.tex
\documentclass[
aps,%
11pt,%
final,%
notitlepage,%
oneside,%
onecolumn,%
nobibnotes,%
nofootinbib,%
superscriptaddress,%
noshowpacs,%
centertags]%
{revtex4}
\input{sao_cmd_author.tex}

\renewcommand{\baselinestretch}{1.3}
\begin{document}
\selectlanguage{english}

\title{H$\boldsymbol{\alpha}$ Images of Ultra-Flat Edge-On Spiral Galaxies}

\author{\firstname{S.~S.}~\surname{Kaisin}} \email{skai@sao.ru} \affiliation{\saoname}

\author{\firstname{I.~D.}~\surname{Karachentsev}} \affiliation{\saoname}

\author{\firstname{H.}~\surname{Hernandez-Toledo}}
\affiliation{Institute of Astronomy (IA) of the UNAM, Mexico City,
04510 Mexico}

\author{\firstname{L.}~\surname{Gutierrez}}
\affiliation{National Astronomical Observatory, Ensenada, 22800
Mexico}

\author{\firstname{V. E.}~\surname{Karachentseva}}
\affiliation{Main Astronomical Observatory NAS of Ukraine, Kyiv, 03143 Ukraine}

\received{September 24, 2019} \revised{November 14, 2019}
\accepted{November 14, 2019}

\begin{abstract}
{We present the H$\alpha$ images of ultra-flat (UF) spiral
galaxies seen practically edge-on. The galaxies have the angular
diameter in the $B$ band \mbox{$a> 1\farcm9$} and  the apparent
axial ratio $(a/b) >10$. We found that their H$\alpha$ images
look, on average, almost two times thinner than those in the red
continuum. The star-formation rate in the studied objects,
determined from the H$\alpha$ flux, is in good agreement with that
calculated from the $FUV$ flux from the GALEX survey if we use the
modified Verheijen and Sancisi formula taking into account the
internal extinction in the UF galaxies. The logarithm of the
specific star-formation rate in the UF galaxies shows a small
scatter, $0.19$, with a smooth decrease from $-10.4$ for dwarf
spirals to $-10.7$ for massive ones. The relative amount of
the hydrogen mass in UF disks varies from about 50\% in dwarf disks to
about 8\% in massive ones. Structural distortions are less common
in the UF galaxies (about 16\%) than those in thick (less isolated)
disks of edge-on spiral galaxies. On the cosmic time scale, 13.7
Gyr, large spiral disks are more efficient ``engines'' for gas
processing into stars than dwarf spirals.} \end{abstract}

\maketitle

\section{INTRODUCTION}
The revised catalog of flat galaxies RFGC contains 4236 objects
distributed over the whole sky (Karachentsev et al. 1999). For the RFGC
catalog, we selected the galaxies with the angular diameter
$a\geq0\farcm6$ in the $B$ band and the apparent axial ratio
$a/b\geq7$ measured in the First Palomar Sky Survey (POSS-1) and
in the ESO/SERC survey. The RFGC catalog includes spiral galaxies
of various morphological types: from S0, Sa to Sd, Sm. The sample
containing 817 ultra-flat galaxies was compiled from this array
(UFgg) (Karachentseva et al. 2016) with the ``blue'' and ``red'' axial
ratios: \mbox{$(a/b)_B\geq10$} and \mbox{$(a/b)_R\geq8.5$}. A
substantial proportion in the UF sample is made up of spiral
galaxies of the Sc, Scd, and Sd types, in which the spheroidal
stellar subsystem makes insignificant contribution to the total
mass/luminosity of a galaxy. Such disk-shaped galaxies with
negligibly small bulges are attractive objects for various studies
of their kinematics, dynamics, and star formation due to the
simple structure of these galaxies.

According to the data from Karachentsev et al. (2016), Melnyk et al. (2017), the UF
galaxies are located in regions of low density avoiding close
proximity to other galaxies. Obviously, the absence of close
neighbors is an important condition for a thin stellar disk to
persist. The presence of very few small satellites in the UF galaxies
makes it possible to estimate the total mass from measurements of
the difference in radial velocities and projected separations of
the satellites. In spite of some expectations of Banerjee and Jog (2013),
dark halos of the UF galaxies did not show any excess of dark
matter compared to other spiral galaxies (Karachentsev et al. 2016).

Obtaining images of ultra-flat galaxies in the H$\alpha$ emission
line is of great interest, since it allows one to distinguish
H\,II regions with young stellar populations. Unfortunately, the
data on the H$\alpha$ images of thin disks of edge-on galaxies are
extremely rare in the literature. However, the first H$\alpha$
image of the UF galaxy RFGC~2246\,$=$\,UGC~7321 has already shown
(Karachentsev et al. 2015) that the subsystem of the young population of the
galaxy has the axial ratio $(a/b)_{{\rm H}\alpha}=38$ which is
much greater than that of the old disk population,
\mbox{$(a/b)=14$}. The relationship between disk flatness and its
population age could be traced from images of the UF galaxies in
the $FUV$ and $NUV$ ultraviolet bands, although, the low angular
resolution of the GALEX survey (Gil de Paz et al. 2007) impedes the success
of this approach.

To observe in the H$\alpha$ line, we selected the largest UF
galaxies with the ``blue'' angular diameter \mbox{$a_B>1\farcm9$}
located in the region of declinations of DEC $>-30^{\circ}$.

\section{OBSERVATIONS AND DATA REDUCTION}
Observations of the UF galaxies in the H$\alpha$ emission line
were carried out at the 2.12-m f/7.5 Casse\-grain telescope of the
San Pedro M\'artir National Astronomical Observatory in Mexico for
several sets since February 2016 till September 2017. The
telescope was equipped with the 2K$\times$2K CCD camera having a
pixel size of 13.5$\times$13.5~$\mu$m. With a binning of
\mbox{2$\times$2}, the camera provided a field of view of
6\arcmin$\times$6\arcmin\  and a resolution of 0.352~arcsec/pixel.
The observations were carried out with a set of narrow
interference filters with equivalent widths of~80\,\AA\ centered
at different wavelengths: 6603, 6643, 6683, and 6723\,\AA\
according to the radial velocity of the galaxy. Figure 1
shows the curves of the spectral transmission of filters.

To subtract the continuum, the images of galaxies were taken in
the r-Gunn broadband filter. The image calibration was carried out
every night using spectrophotometric standards.

The reduction of the observed data was carried out using the set of
standard procedures which included: bias subtraction, flat-field
division, cosmic-ray removal, and sky-background subtraction. The
images in the continuum were normalized to the image in the
H$\alpha$ filter using several dozen stars and then were
subtracted. The H$\alpha$ flux of the galaxy was determined from
the H$\alpha$ image with the subtracted continuum. The typical
measurement error of the H$\alpha$ flux was determined, as a rule,
by weather conditions and it was equal to about~$0.1$~dex. With this
accuracy, we ignored the contribution of the~[N\,II] doublet to the emission
flux neighboring H$\alpha$.

\section{RESULTS}

Figure~2  gives the first page of the mosaic comprising 45 pairs
of the UF-galaxy images that we obtained. The left-hand images in
each pair correspond to the total exposure in the H$\alpha$ line
and in the continuum, and the right-hand ones show the image
difference in H$\alpha$ and in the continuum. The name of each
object, the scale of the image, and the ``north--east''
orientation are indicated in the right-hand images. Some images
show residual traces from bright stars and objects of an
anomalous color.

Table~1 summarises the main parameters of the observed
UF galaxies. The table columns contain: (1) the number of the
galaxy in the RFGC catalog; (2) the equatorial coordinates; (3,4)
the ``blue'' angular diameter in arcmin and apparent ``blue''
axial ratio from RFGC; (5) the heliocentric radial velocity of the
galaxy in km\,s$^{-1}$; (6) the effective wavelength of the filter
(\AA) in which the galaxy was exposed; (7) the exposure time in
the H$\alpha$ filter in sec; (8) the flux logarithm in the
H$\alpha+$\,[N\,II] lines in erg\,cm$^{-2}$~\,s$^{-1}$.

Besides the H$\alpha$ line, the [N\,II] nitrogen lines fall into
the filters we used: 6548\,\AA\, and 6584\,\AA. According to
Kennicutt et al. (2008), the relation of the intensities of the [N\,II]
and H$\alpha$ lines for spiral galaxies depends on the absolute
magnitude of the galaxy and is expressed by the relation
\begin{equation} \label{eq1} \log(F{\rm [N\,II]}/F({\rm
H}\alpha))=-0.173 M_B -3.90 \end{equation}
with $M_B>-21\fm0$ and
$-0.27$, when $M_B<-21\fm0$, having a standard deviation of
$0.26$~dex. For a typical galaxy of our sample with
\mbox{$M_B\simeq -19\fm4$}, the correction to the flux $F({\rm
H\alpha})$ due to the contribution of the [N\,II] doublet is
$-0.14$~dex, which is smaller than the standard deviation in
relation~(1). That is why, we did not correct the measured
flux \mbox{$F({\rm H}\alpha+{\rm [N\,II]})$} for the contribution
of the nitrogen doublet.

We used the measured integrated flux of the galaxy $F_c({\rm
H}\alpha)$ corrected for Galactic and internal extinction to
determine the star-formation integral rate, $SFR({\rm H}\alpha$),
in units $M_{\odot}$/yr. According to Kennicutt et al. (1998),
\begin{equation} \log(SFR({\rm H}\alpha))=\log F_c({\rm
H}\alpha)+2\log D +8.98, \end{equation} where the distance $D$ is
given in Mpc.

Most galaxies in our sample have the estimated apparent magnitudes
$m_{FUV}$ in the $FUV$ band of the far ultraviolet ($\lambda_{\rm
ef}=1539$\,\AA, FWHM\,$=269$\,\AA) measured at the
GALEX\footnote{\url{http://galex.stsci.edu/GalexView/}} satellite.
Following Lee et al. (2011), we determined the integral
star-formation rate of the galaxy as \begin{equation}
\log(SFR(FUV))= 2.78-0.4 m^c_{FUV}+2\log D, \end{equation} where
the apparent $FUV$ magnitude is corrected for the Galactic and
internal extinction. Comparison between $SFR({\rm H}\alpha$) and
$SFR(FUV)$ makes it possible to refine the value of internal
extinction in galaxies, which appears to be significant in the
case of edge-on galaxies.

Table~2 presents the extended summary of the main
parameters of the UF galaxies. Besides those 45 galaxies that we
observed, we also included the data on 10 UF galaxies with
the $F({\rm H}\alpha)$ flux measurements conducted in Karachentsev et al. (2015),
Gavazzi et al. (2015), Spector and Brosch (2017)
at the end of the table. The
columns of Table~2 contain: (1) the RFGC galaxy number; (2) the  morphological type according to the de
Vaucouleurs classification: 4---Sbc, 5---Sc, 6---Scd, 7---Sd that
we determined from the galaxy images in the PanSTARRS survey
(Chambers et al. 2016); (3) the logarithm of the apparent axial ratio
reduced to the standard isophote from HyperLEDA (Makarov et al. 2014);
(4, 5) the apparent $B$ magnitude and the Galactic extinction in the
$B$ band from Makarov et al. (2014), Schlegel et al. (1998); (6) the distance to the
galaxy (Mpc) determined from the radial velocity relative to the
Local Group centroid with the Hubble parameter
$H_0=73$~km\,s$^{-1}$\,Mpc$^{-1}$; for closer galaxies with
$V_{\rm LG} <2500$~km\,s$^{-1}$, the estimation of $D$ is
made using the Shaya et al. (2017) model which takes into account the
infall of galaxies to the Virgo cluster and expansion of the Local
cosmic void; (7) the amplitude of the galaxy rotation
(in~km\,s$^{-1}$) from Makarov et al. (2014); (8) the apparent
magnitude $m_{21}$ from  Makarov et al. (2014) characterizing the flux
from the galaxy in the 21-cm line of neutral hydrogen; (9) the
flux logarithm in the H$\alpha$ line; (10) the apparent
magnitude of the galaxy in the $FUV$ band from the GALEX data;
(11) the logarithm of the hydrogen mass of the galaxy
\begin{equation} \log M_{\rm HI}=12.33 -0.4 m_{21}+2\log D
\end{equation} expressed in units $M_{\odot}$; (12) the
magnitude of the accepted internal extinction in the galaxy in
the $B$ band (see Section~4); (13) the apparent
magnitude of the galaxy in the $K$ band determined from the integral $B$ value
and from the morphological type $T$ as \begin{equation} K = B +T/4
-4.60
\end{equation} corrected for the Galactic and internal
extinction; such a recipe suggested in Jarrett et al. (2003) eliminates
systematic underestimation of the flux from peripheral regions
under photometry of blue edge-on galaxies in the 2MASS survey
(Jarrett et al. 2000); (14) the integral luminosity of the galaxy in the
$K$ band (in $L_{\odot}$) which with \mbox{$M_*/L_K=1
M_{\odot}/L_{\odot}$} (Bell et al. 2003) corresponds to the stellar
mass of the galaxy; (15,16) the integral star-formation rate
determined from the H$\alpha$ and $FUV$ fluxes, respectively; (17)
the specific star-formation rate $sSFR({\rm H}\alpha)/M_*$ in
units (yr$^{-1}$) under the assumption that \mbox{$M_*/L_K=1$} in
solar units.

\section{ACCOUNTING FOR THE INTERNAL EXTINCTION IN THE UF GALAXIES}\label{Sect4}

The example of our Galaxy shows that dust, H\,II regions, and blue
stars are distributed in a spiral disk extremely unevenly. The
picture of the shredded distribution of dust is far from a simple model of
flat-parallel layers. For this reason, a reliable scheme for
accounting for internal extinction has not yet been proposed.
Usually, the extinction in the $B$ band is expressed as
\begin{equation} A^i_B= \gamma \log(a/b), \end{equation} where
the coefficient $\gamma$ depends on the luminosity or
morphological type of the galaxy. The HyperLEDA accounting scheme
for the internal extinction implies a dependence of $\gamma$ on a
morphological type. Its imperfection is the monotonic increase of
$\gamma$ with the increase of $T$ which leads to strong
overestimation of the extinction in late-type dwarf galaxies.

Other authors (Bothwell et al. 2009, Devour and Bell 2016, Lee et al. 2009)
used the schemes,
where the parameter $\gamma$ depended on the absolute magnitude of
the galaxy, and the character of this dependence was significantly
different for different authors. Obviously, the absolute magnitude
of the galaxy itself depends on the accepted internal extinction,
so, the $A^i_B$ estimation scheme for edge-on galaxies
requires a series of sequential iterations.

Verheijen and Sancisi (2001) proposed to express the parameter $\gamma$ via the
amplitude of the galaxy rotation: \begin{equation}
\gamma(V_m)=1.54+2.5(\log V_m-2.2) \end{equation} with $V_m>43$
km\,s$^{-1}$, otherwise $\gamma=0$, when 
$V_m<43$~km\,s$^{-1}$. This approach is free from
iterations, however, it is applicable only to galaxies with a
known rotation amplitude. Considering the statistics of the
relation $SFR(H\alpha)/SFR(FUV)$ for the Local Volume galaxies,
Karachentsev et al. (2018) concluded that expression (6) somewhat
overestimates the extinction in massive galaxies and
underestimates it for dwarf galaxies. In our estimation, the
appropriate correction for the internal extinction in late-type
spiral disks has the form \begin{equation} A^i_B=(1.3+2.0 (\log
V_m-2.2))\log r_{25} \end{equation} with $V_m>36$~km\,s$^{-1}$,
otherwise $A^i_B=0$, when \mbox{$V_m<36$}~km\,s$^{-1}$, where
$r_{25}=(a/b)_{25}$~is the apparent axial ratio reduced to the
standard isophote (Makarov et al. 2014). Table~2 presents
$A^i_B$ calculated according to this recipe. For two galaxies with
unknown $V_m$, we estimated the extinction with the empirical
relation:
\begin{equation} A^i_B(T) = \left\{\begin{array}{c}(3.0-0.3 T)
\log r_{25}, T>4;\\ 0.3 (1+T) \log r_{25}, T<5, \end{array}
\right.
\end{equation} which describes extinction in late-type galaxies
more adequately than the schemes from Bothwell et al. (2009), Devour and Bell (2016),
Lee et al. (2009), or the algorithm used in
HyperLEDA. Following Lee et al. (2009), we accepted the transition
coefficients for the Galactic extinction in the $H\alpha$ and
$FUV$ bands:
\begin{equation} A^G_{{\rm H}\alpha} = 0.61 A^G_B,\,\,\,
A^G_{FUV}=1.93 A^G_B. \end{equation} For the internal extinction,
according to Lee et al. (2009), these relations were accepted:
\begin{equation} A^i_{{\rm H}\alpha}=1.07 A^i_B,
\,\,\,A^i_{FUV}=1.93 A^i_B.
\end{equation} Here, a higher value of the transition coefficient
for the H$\alpha$ line compared to relation (10) is due to the
close correlation between the distribution of dust and H\,II
regions in the galaxy disks, and its value was estimated from
spectrophotometric measurements of the Balmer decrement (see
the details in Lee et al. 2009).

Determining the hydrogen mass of galaxies $M_{\rm HI}$, we ignored
the correction for the internal self-extinction of the emission
in the 21-cm line. For the edge-on galaxies, HyperLEDA introduces a
correction to $m_{21}$ for the self-extinction effect, equal to
\mbox{$\Delta m_{21}=-0\fm82$}. However, such a correction seems
overstated to us.  Jones et al. (2018) investigated the self-extinction
effect in the 21-cm line for the sample of 2022 galaxies from the
ALFALFA survey and concluded that galaxy disks are almost
transparent in the 21-cm line, and the required correction is only
\begin{equation} \Delta\log M_{\rm HI}= (0.13\pm0.03) \log (a/b).
\end{equation}

The comparison between the UF and Sc, Sd face-on galaxy samples
(Karachentsev and Karachentseva 2019) shows that the self-extinction effect is actually
even smaller being lost in measurement errors of the H\,I flux of
galaxies and errors in the morphological classification of
galaxies.

\section{STAR-FORMATION RATES IN ULTRA-FLAT GALAXIES}
The upper panel of Figure~3 shows the dependence between
the star-formation rate determined from the H$\alpha$ flux and the
$K$ luminosity of the UF galaxies. Our measurements are shown with
the solid circles and data from the literature---with the open
circles. The dashed line corresponds to the case
\mbox{$\log(SFR)=\log L_K-10.14$}, when the galaxy manages to
reproduce its observed stellar mass with the observed $SFR$ for
the cosmological time \mbox{$T_0=13.7$}~Gyr. The linear regression
(the solid line) has a slope of $0.87\pm0.04$ indicating that more
massive galaxies required higher star-formation rates in the past
to provide the accumulated stellar mass. The similar diagram in
the case of $SFR$ calculated from the $FUV$ flux is shown in the
lower panel of Fig.~3. In general, the diagram has a
similar shape, although, the dispersion of the $SFR$ estimates is
larger in Fig. 3b.

Figure~4 gives the comparison of the obtained
$SFR(H\alpha)$ and $SFR(FUV)$ values. The data is well grouped
along diagonal, having average\linebreak \mbox{$\langle
SFR(H\alpha)\rangle =-0.11\pm0.08$} and $\langle SFR(FUV)\rangle
=-0.03\pm0.09$. This circumstance indirectly confirms that the
difference in calibrations of empirical relations (1) and (2) is
small, and the scheme we have adopted for taking into account the
internal extinction in galaxies is close to reality.

Figure~5 reproduces the relationship between the specific
star-formation rate $sSFR(H\alpha)$ and the total $K$
luminosity or stellar mass of the UF galaxies with $M_*/L_K=1$ in solar units.
The dashed horizontal line corresponds to the Hubble parameter
\mbox{$H_0=73$}~km\,s$^{-1}$\,Mpc$^{-1}$. The scatter of galaxies
relative to the quadratic regression line is small, $0.19$~dex,
which indicates a fairly uniform pattern of the star formation in thin disks of
late-type spiral galaxies. Moreover, in massive disks, the
gas-to-star conversion occurred in the past at about two
times higher rates than in dwarf spirals.

It should be noted that this difference is leveled if $SFR$ is
normalized not to the stellar mass but to the total baryon mass of
the galaxy.

\section{SOME MAIN PROPERTIES OF ULTRA-FLAT SPIRAL DISKS}
The ultra-flat category represents the galaxies with a large range
of linear sizes. The nearby Sd dwarf, RFGC\,1700 $=$
UGCA\,193 has the minimum linear diameter in our sample, 13 kpc.
Among the giant disks, the Sbc galaxy RFGC\,1339 $=$ UGC\,4704 has
the largest diameter, 105 kpc. The median linear diameter of the
UF galaxies is 44 kpc. The correction for inclination adopted in HyperLEDA about
one and a half times decreases the isophotal diameter of the UF
galaxy.

Figure~6 shows the dependence between the hydrogen mass
and the $K$ luminosity of the UF galaxies. As follows from these
data, the ratio $M_{\rm HI}/L_K$ systematically decreases from
dwarf galaxies to high-luminosity objects. This pattern indicates
that the process of converting gas into stars was the most intense
in the most massive galaxies. The noticed effect obviously is not
related to the presence of bulges in galaxies, since their
contribution to the luminosity of the UF galaxies is quite small.

Comparing the logarithm of the hydrogen mass for the face-on
(Karachentsev and Karachentseva 2019) and edge-on galaxies within equal intervals of the
$L_K$ luminosities, we got the average difference
$\langle\log M_{\rm HI}\rangle_{\rm faceon}-\langle\log
M_{\rm HI}\rangle_{\rm edgeon}=-0.08\pm0.06$. The negative value
of this difference under the typical ratio $\log(a/b)\simeq1$ for
the UF galaxies indicates that the disks of ultra-flat galaxies
are almost transparent in the 21-cm line, and correction (12) for
them is excessive.

The upper and lower panels of Fig.~7 show the dependence
of the star-formation rate determined from the H$\alpha$ and $FUV$
fluxes on the hydrogen mass of the UF galaxies. The regression
lines on them have an inclination of $1.27\pm0.12$ and
$1.16\pm0.08$, noticeably smaller than the expected $1.4\pm0.1$
from the Schmidt--Kennicutt relation (Kennicutt 1998) for
individual star-formation sites. It should also be noted that
the dispersion of the observed data on the $SFR$--$M_{\rm HI}$
diagrams obtained from the $FUV$ fluxes is noticeably smaller
than that obtained from the H$\alpha$ fluxes unlike the \mbox{$SFR$--$L_K$}
diagrams (Fig.~3). We did not find an explanation for
this feature.

Comparison of the images given in Fig.~2 shows that all
the UF galaxies without exception appear thinner in the H$\alpha$
filter than in the red continuum. This difference is the stronger,
the closer the inclination angle of the galaxy to $i = 90^{\circ}$
is. Figure~8 reproduces the ratio~$a/b$ in the H$\alpha$
line and in the red continuum for 45 galaxies under study. Average
values of \mbox{$\langle\log(a/b)_{{\rm
H}\alpha}\rangle=1.23\pm0.03$} and
\mbox{$\langle\log(a/b)_r\rangle=0.97\pm0.02$} show that the
thickness of the emission disk is on average almost two times
smaller than that in the red continuum. As is known, the complexes
of hot blue stars that regulate the glow of H\,II regions are
approximately~$10^7$ yrs old. Consequently, the formation of a
young stellar population occurs in a thinner layer of the disk
compared to the thickness of the disk of an old stellar
population. This conclusion is quite expected in the picture of
the formation of young H\,II complexes with gravitational
instability of molecular gaseous clouds.

Reshetnikov and  Combes (1998) investigated the statistics of S-like
distortions in optical images of flat galaxies. According to the data by
Reshetnikov and  Combes (1998), such
distortions are visible in 40\% of edge-on galaxies, and their
frequency increases with the increase of the flat galaxy environment. The latter circumstance indicates the
external, tidal nature of the distortions visible on the outskirts
of the disks. In our sample of 45 UF galaxies, we found distinct
distortions of the emission disk only in one galaxy, RFGC\,1133
$=$ UGC\,3539\footnote{This integral-shaped galaxy is highly
isolated. Its nearest neighbor, the galaxy CGCG\,308-039, has a
radial velocity difference of~$228$~km\,s$^{-1}$ and a projected
separation of 410~kpc.}, and weak distortions for other 6 galaxies:
RFGC\,504, 531, 722, 1434, 3935, and 4039. Thus, the occurrence of
distortions of the H$\alpha$ disk of ultra-flat galaxies, no
greater than $(16\pm5)$\%, turns out to be noticeably smaller than that of
the objects of the RFGC catalog. A small percentage of peripheral
distortions in ultra-flat galaxy disks is in agreement with the
fact that they are preferred in the very low density regions.

\section{FINAL REMARKS}
The presented results of observations in the H$\alpha$ line of
ultra-flat galaxies resulted in a multiple increase in the number
of the studied objects of this category. The UF edge-on galaxies
have angles of rotational axis inclination to the line of sight in
the range of $i\simeq(85$--$90)^{\circ}$ which, with the absence
of significant bulges, corresponds to the apparent axial ratio
\mbox{$a/b>10$} in the blue region of the spectrum. In the
emission H$\alpha$ line, the UF galaxies look even thinner
having the characteristic axial ratio \mbox{$\langle
a/b\rangle_{{\rm H}\alpha}\simeq17$}. This shows that the
young stellar population of galaxy disks is formed in a narrow
layer, the thickness of which increases with the transition to an
older population.

The internal extinction in the UF galaxies appears to be significant. With a
characteristic linear diameter of about~$44$~kpc, the extinction
in the H$\alpha$ line reaches \mbox{$1$--$2^{\rm m}$}, and in the
$FUV$ band---even $3$--$4^{\rm m}$. The considerable extinction
results in the faint appearance of the UF galaxies in the GALEX
ultraviolet sky survey. The method of accounting for the
internal extinction used by us leads to good agreement between
the estimates of the star-formation rate obtained from the
H$\alpha$ and $FUV$ fluxes. In the 21-cm emission line, the
ultra-flat galaxies of our sample are almost transparent.

The specific star-formation rate in the UF galaxies, referred to
the unit of the $K$ luminosity or stellar mass, shows a systematic
decrease from \mbox{$sSFR\sim -10.4$} dex with $L_K\sim9$ dex to
about \linebreak$-10.7$~dex with \mbox{$L_K\sim11$}~dex. The low
dispersion on the $sSFR$ vs $L_K$ diagram relative to the
regression line indicates the uniformity of the star formation in
thin disks of spiral galaxies.

To reproduce the observed stellar mass, the average star-formation
rate of dwarf and massive UF galaxies had to be two and four times
higher in the past, respectively, than their current value of
$sSFR$.

The relative abundance of hydrogen mass in the UF galaxy disks is
on average about 20\%, varying from 50\% in dwarf disks to about
8\% in massive galaxies. Consequently, the UF galaxies have gas
reserves to maintain the observed star-formation rates for over several
billion years.

Disk shape distortions are noticeably less common in the UF
galaxies than those in other edge-on galaxies of the RFGC catalog. The
presence of the UF galaxies in the regions of low cosmic density
is consistent with the assumption that many distortions of the
periphery of spiral galaxies are due to the tidal force of
close neighbors.

\begin{acknowledgements}
In this paper, we used the data of the GALEX and PanSTARRS sky
surveys as well as the HyperLEDA extragalactic database.
This work was partially supported by the RSF grant 19--12--00145.
\end{acknowledgements}

\clearpage

\begin{figure} \setcaptionmargin{5mm}
\onelinecaptionstrue \captionstyle{normal}
\includegraphics[scale=1.2]{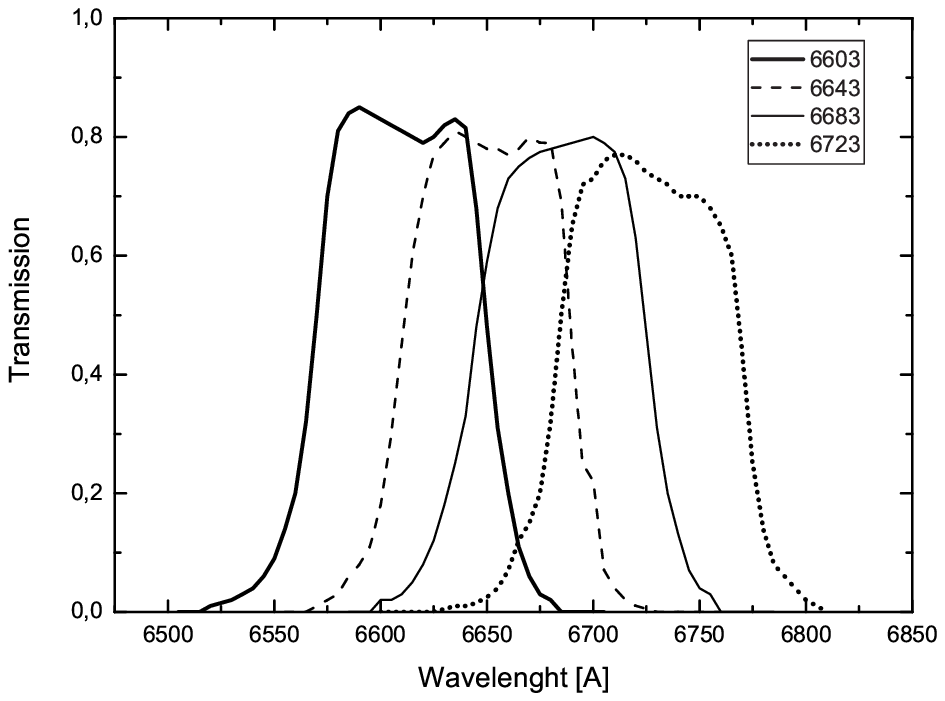}
\caption{Curves of the spectral transmission of filters used in
observations.} \label{fig1} \end{figure}

\begin{figure*}
\setcaptionmargin{5mm} \onelinecaptionstrue \captionstyle{normal}
\includegraphics[scale=0.2]{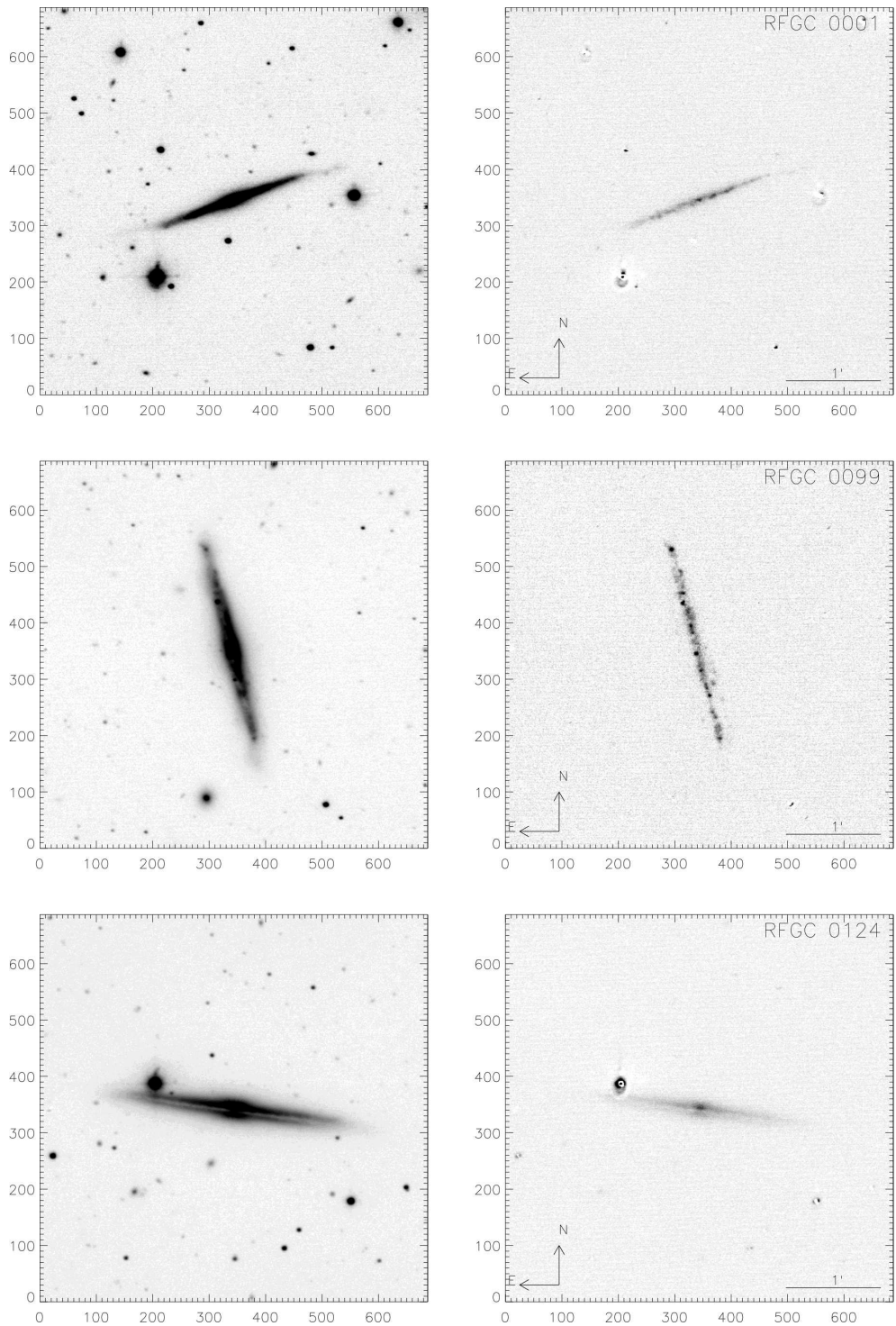} \caption{Mosaic
image of ultra-flat galaxies. The left-hand images in each pair
present the sum of the exposures in the H$\alpha$ line and in the
continuum, and the right-hand images correspond to the difference
``H$\alpha$--continuum''. The right-hand images show: the name of
the galaxy, $1\arcmin$ scale, and``North--East'' direction. The
full consolidated data on the H$\alpha$ images of the UF galaxies
are available at {\tt http://lv.sao.ru/EDGE-ON/}.} \label{fig2}
\end{figure*}

\begin{figure} \setcaptionmargin{5mm}
\onelinecaptionstrue \captionstyle{normal}
\includegraphics[scale=0.2]{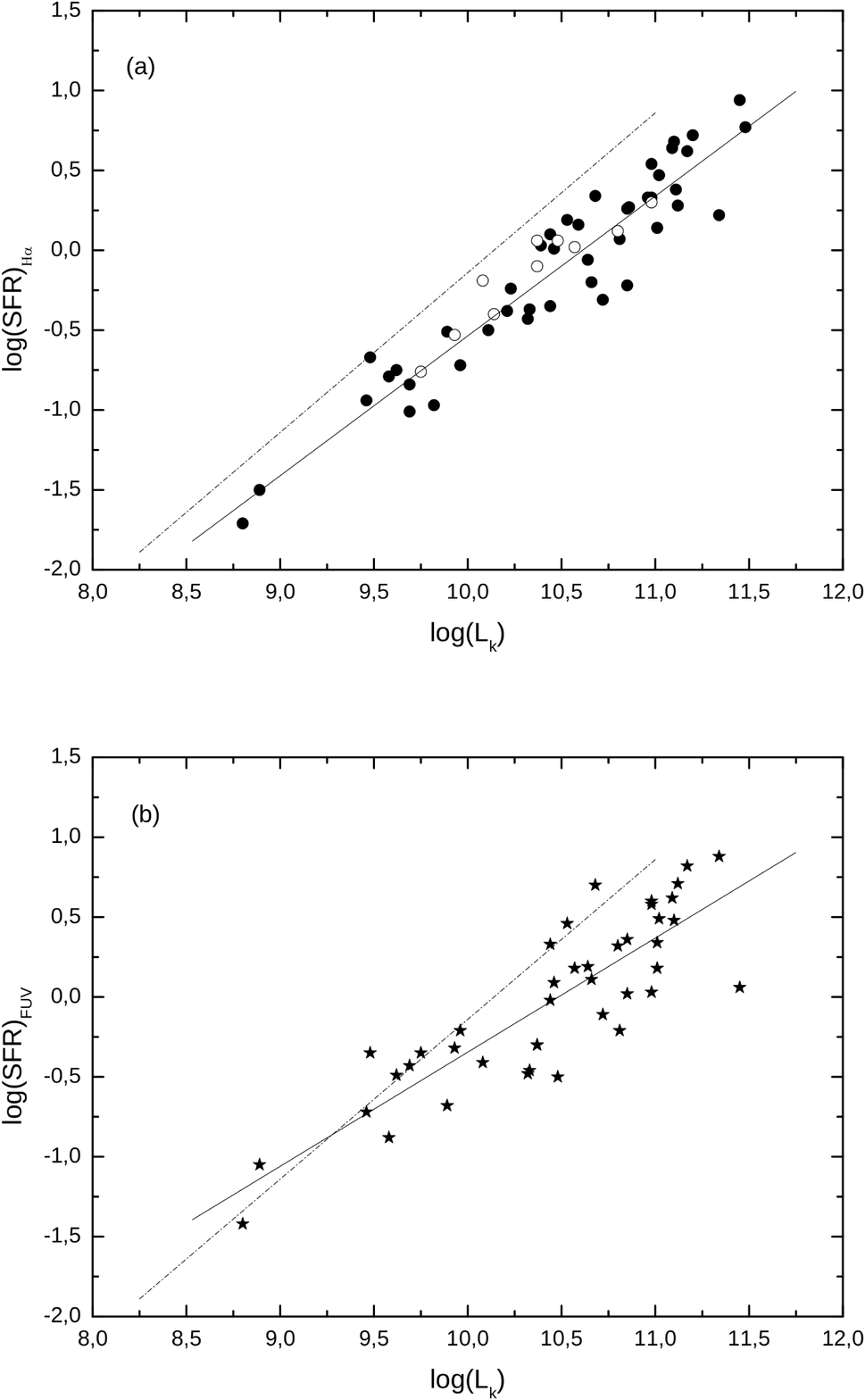}
\caption{Dependence of the star-formation rate determined from a)
the H$\alpha$ flux, b) the flux in the $FUV$ band on the $K$
luminosity of the galaxies. The data of the H$\alpha$ fluxes from
the literature are denoted by the open circles. The dashed line
corresponds to a cosmic time of $13.7$~Gyr, for which the observed
stellar mass of the galaxy is reproduced at the observed $SFR$
rate. The linear regressions have an inclination of $0.87\pm0.04$
and $0.71\pm0.06$ for the H$\alpha$ and $FUV$ fluxes,
respectively.} \label{fig3} \end{figure}

\begin{figure} \setcaptionmargin{5mm}
\onelinecaptionstrue \captionstyle{normal}
\includegraphics[scale=0.12]{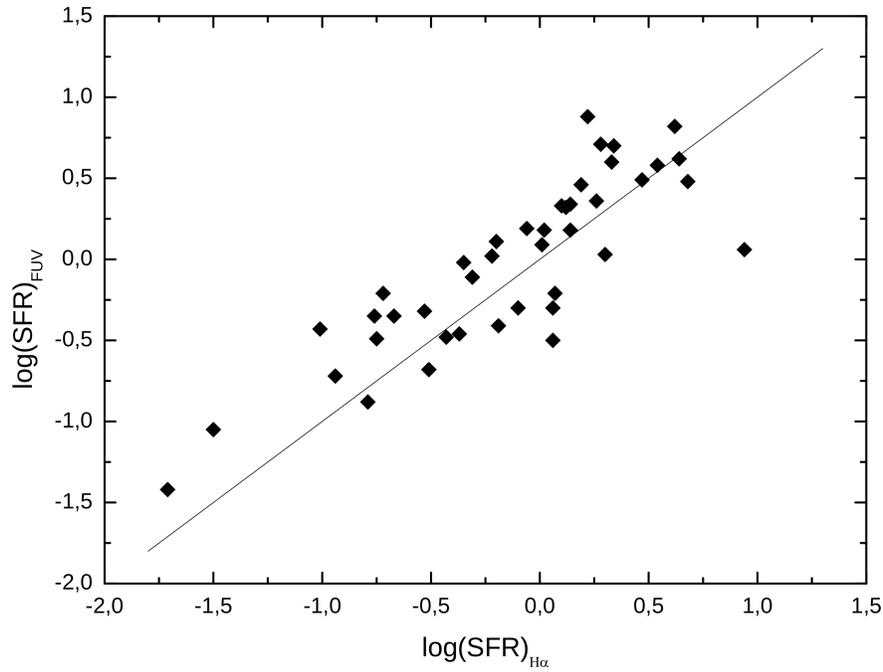} \caption{Relation
between the $SFR$ estimates obtained from the H$\alpha$ and $FUV$
fluxes for the UF galaxies.} \label{fig4} \end{figure}

\begin{figure} \setcaptionmargin{5mm}
\onelinecaptionstrue \captionstyle{normal}
\includegraphics[scale=0.12]{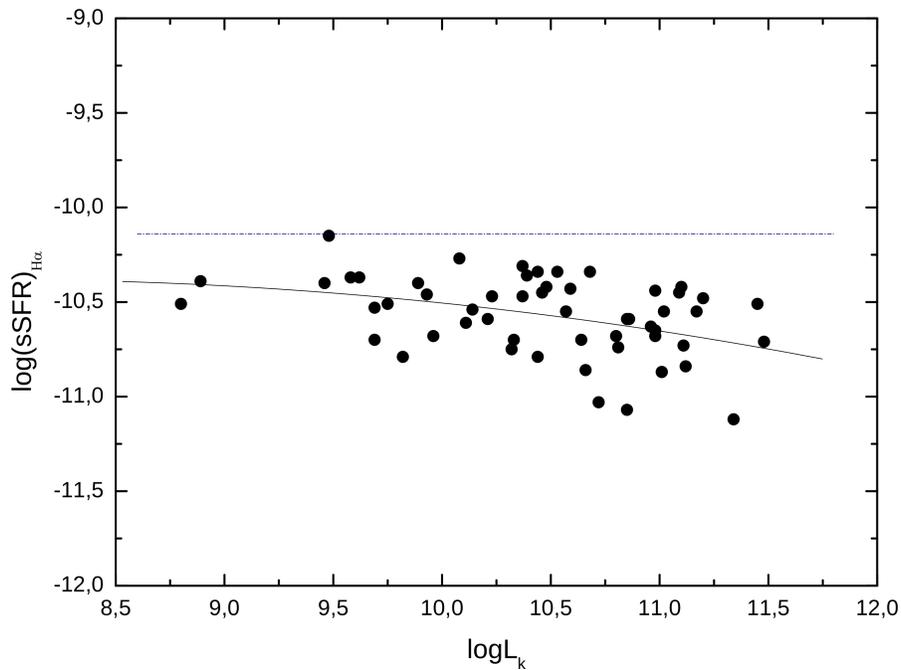} \caption{Dependence
of the specific star-formation rate on the $K$ luminosity for the
galaxies observed in the H$\alpha$ line. The dashed horizontal
line corresponds to the Hubble parameter $H_0 =
73$~km\,s$^{-1}$\,Mpc$^{-1}$. The solid line indicates the
quadratic regression.} \label{fig5} \end{figure}

\begin{figure} \setcaptionmargin{5mm}
\onelinecaptionstrue \captionstyle{normal}
\includegraphics[scale=0.12]{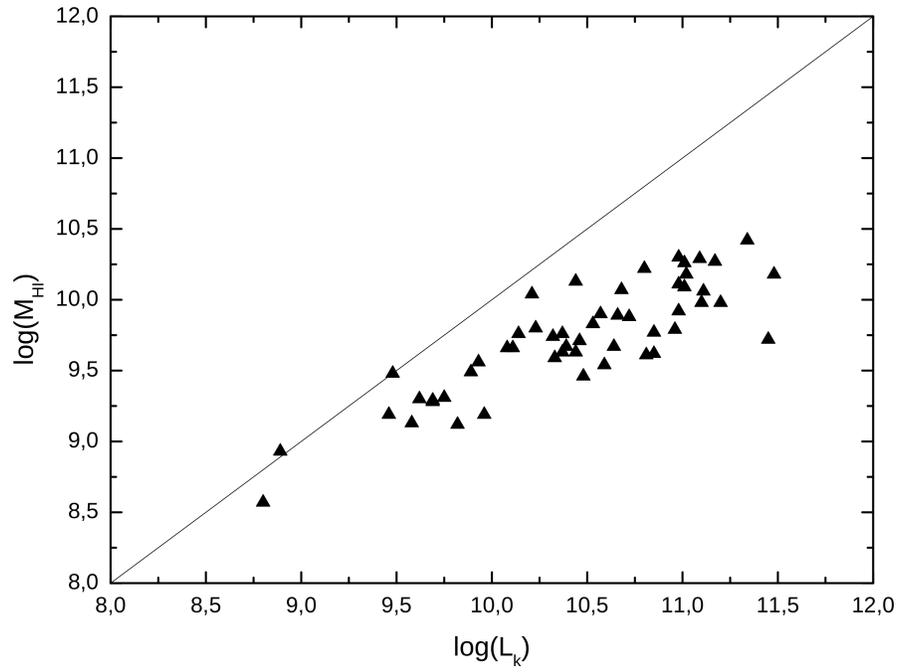} \caption{
Distribution of ultra-flat galaxies with the hydrogen integral
mass and the $K$ luminosity.} \label{fig6} \end{figure}

\begin{figure} \setcaptionmargin{5mm}
\onelinecaptionstrue \captionstyle{normal}
\includegraphics[scale=0.20]{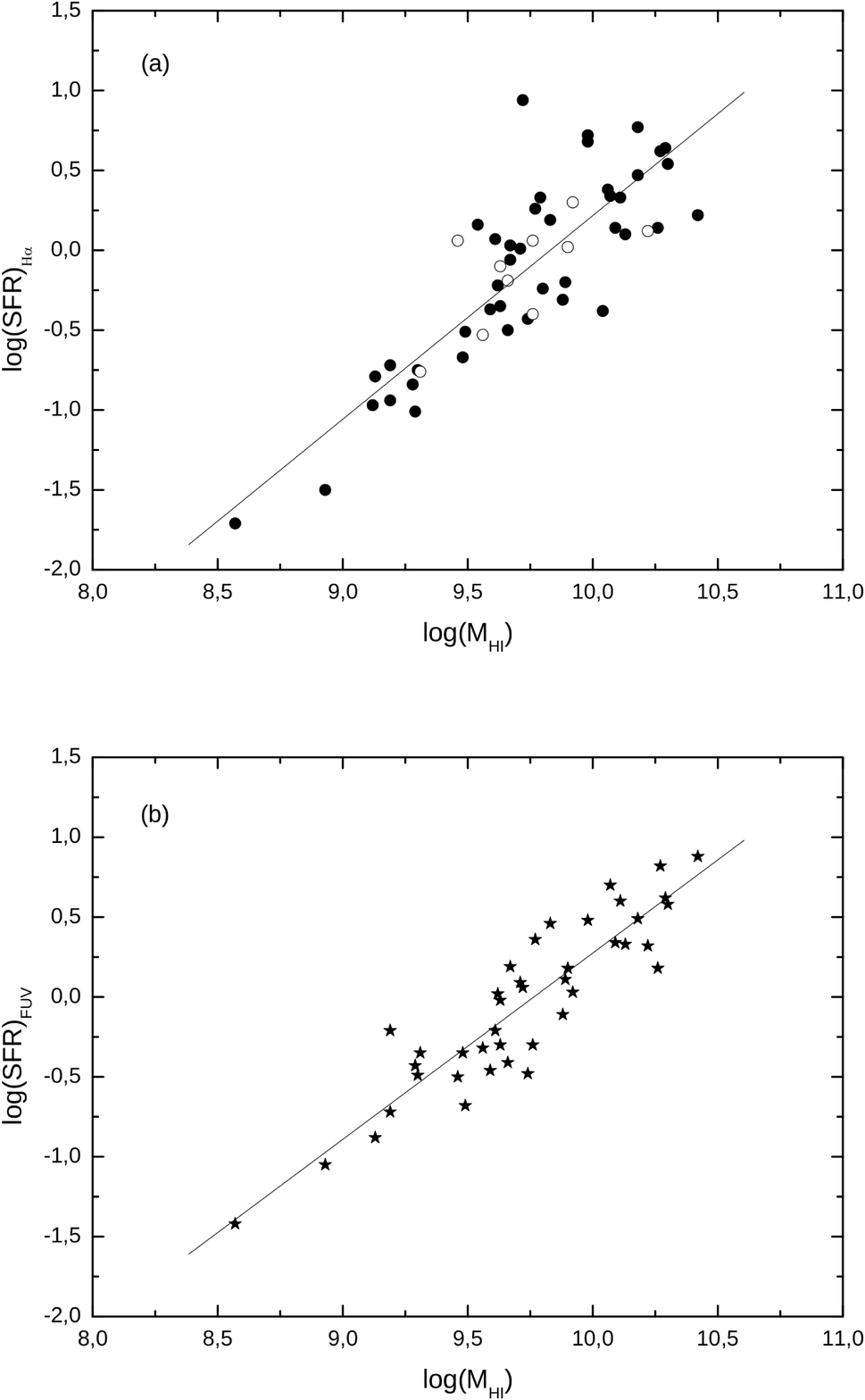}
\caption{Dependence of the star-formation rate determined from a)
the H$\alpha$ flux, b) the flux in the $FUV$ band on the hydrogen
mass. The linear regressions have inclinations of $1.27\pm0.12$
and $1.16\pm0.08$ for the H$\alpha$ and $FUV$ fluxes,
respectively.} \label{fig7} \end{figure}

\begin{figure} \setcaptionmargin{5mm}
\onelinecaptionstrue \captionstyle{normal}
\includegraphics[scale=0.12]{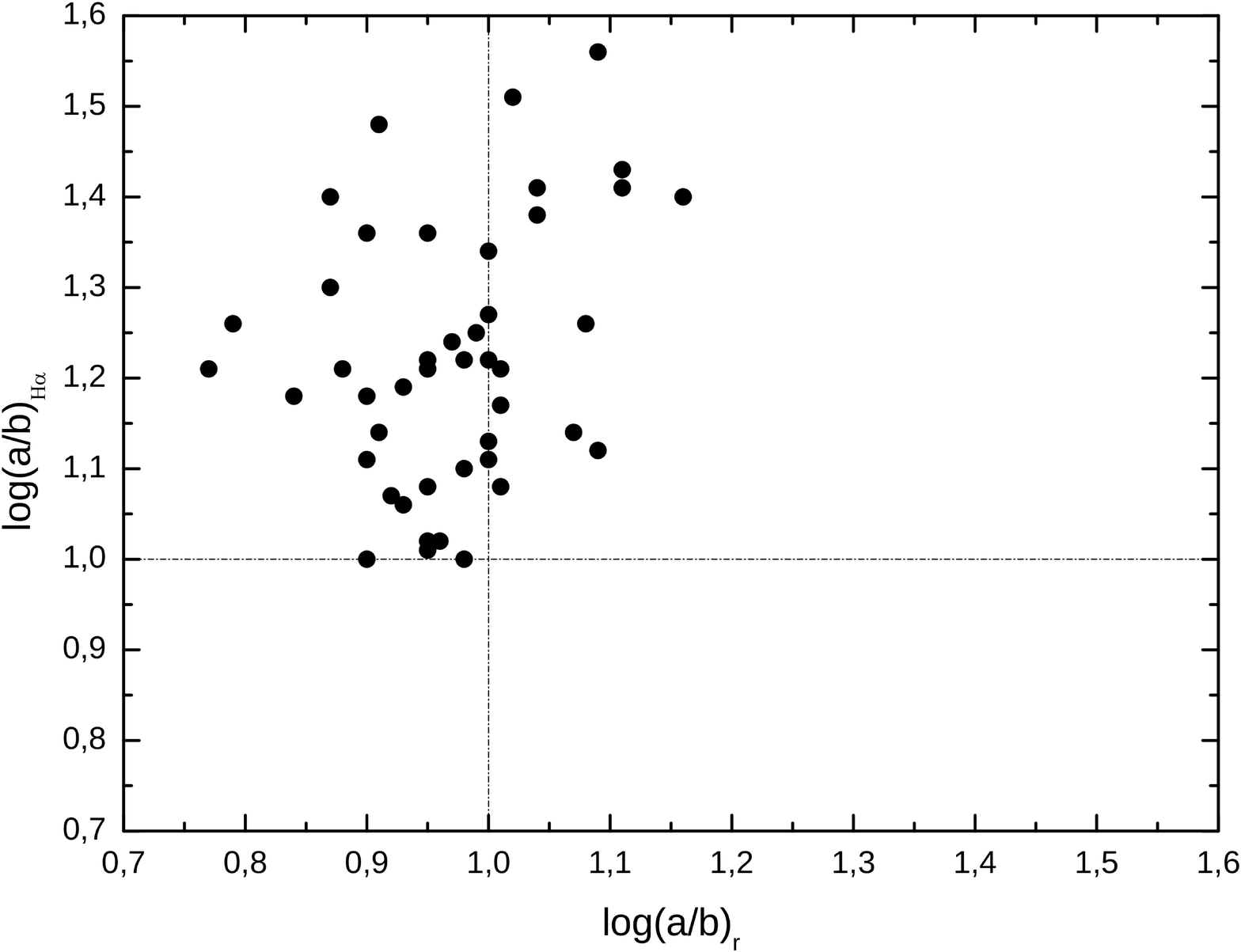} \caption{Apparent
axial ratio in the H$\alpha$ line and in the continuum for the UF
galaxies.} \label{fig8} \end{figure}

\clearpage
\input{Kaisin_tab1.tex}
\clearpage
\input{Kaisin_tab2.tex}

\end{document}

%% file: sao_cmd_author.tex
\def\saoname{Special Astrophysical Observatory,  Russian Academy of Sciences,
              Nizhnii Arkhyz, 369167 Russia}

\def\squareforqed{\hbox{\rlap{$\sqcap$}$\sqcup$}}

\def\sq{\ifmmode\squareforqed\else{\unskip\nobreak\hfil
\penalty50\hskip1em\null\nobreak\hfil\squareforqed
\parfillskip=0pt\finalhyphendemerits=0\endgraf}\fi}

\def\arcmin{\hbox{$^\prime$}}

\def\utw{\smash{\rlap{\lower5pt\hbox{$\sim$}}}}

\def\udtw{\smash{\rlap{\lower6pt\hbox{$\approx$}}}}

\def\fm{\hbox{$\,.\!\!^{\rm m}$}}

\def\farcm{\hbox{$\,.\mkern-4mu^\prime$}}

\def\diameter{{\ifmmode\mathchoice
{\ooalign{\hfil\hbox{$\displaystyle/$}\hfil\crcr
{\hbox{$\displaystyle\mathchar"20D$}}}}
{\ooalign{\hfil\hbox{$\textstyle/$}\hfil\crcr
{\hbox{$\textstyle\mathchar"20D$}}}}
{\ooalign{\hfil\hbox{$\scriptstyle/$}\hfil\crcr
{\hbox{$\scriptstyle\mathchar"20D$}}}}
{\ooalign{\hfil\hbox{$\scriptscriptstyle/$}\hfil\crcr
{\hbox{$\scriptscriptstyle\mathchar"20D$}}}}
\else{\ooalign{\hfil/\hfil\crcr\mathhexbox20D}}%
\fi}}

\newcommand{\ab}{Astrophysical Bulletin }

\newcommand{\aaa}{Astron. and Astrophys. }

\newcommand{\aj}{Astron.~J. }
\newcommand{\apjs}{Astrophys.~J. Suppl. }

\newcommand{\mnras}{Monthly Notices Royal Astron. Soc. }



%% file: Kaisin_tab1.tex
  \renewcommand{\baselinestretch}{0.7}
  \begin{table*}
\setcaptionmargin{0mm} \captionstyle{normal}
 \caption{List of the UF galaxies observed in H$\alpha$ line}
 \label{table1}
 \medskip
 \begin{tabular}{l|c|c|c|r|c|c|c}  \hline
 ~~~~Galaxy  & RA (2000.0) DEC &  a$^{\prime}$ &$ \log(a/b)$& $V_h$~~~ &   Filter & $T_{\rm exp}$    & $\log F_{{\rm H}\alpha}$\\ \hline
 \qquad (1)     &        (2)        &  (3)  &    (4)&  (5)~~ &       (6) &   (7)  &     (8) \\ \hline

RFGC~001 & 000056.0 $+$202016 & 2.02 &  1.07 &  6804 &    6723 &  1440  &   $-$13.45   \\
RFGC~099 & 002547.7 $-$021705 & 2.46 &  1.05 &  5339 &    6643 &  2160  &   $-$12.97   \\
RFGC~124 & 003149.4 $-$264312 & 2.80 &  1.00 &  7235 &    6723 &  2160  &   $-$12.72   \\
RFGC~161 & 004214.7 $-$180942 & 3.36 &  1.05 &  1553 &    6603 &  2700  &   $-$12.80   \\
RFGC~176 & 004708.2 $+$302027 & 2.50 &  1.08 &  5248 &    6643 &  2160  &   $-$12.98   \\
RFGC~255 & 010836.9 $+$013830 & 4.65 &  1.19 &  1982 &    6603 &  2700  &   $-$12.77   \\
RFGC~438 & 020302.0 $-$093922 & 2.80 &  1.19 &  3864 &    6643 &  1800  &   $-$12.73   \\
RFGC~463 & 020926.4 $+$371529 & 2.13 &  1.29 &  4586 &    6643 &  2160  &   $-$13.46   \\
RFGC~504 & 022131.0 $+$141155 & 2.52 &  1.06 &  3744 &    6643 &  2160  &   $-$12.85   \\
RFGC~511 & 022356.1 $-$064216 & 1.97 &  1.22 &  9560 &    6723 &  2160  &   $-$13.61   \\
RFGC~517 & 022515.5 $+$452704 & 2.02 &  1.00 &  5195 &    6643 &  2160  &   $-$13.21   \\
RFGC~531 & 022827.3 $+$153625 & 1.90 &  1.16 &  4080 &    6643 &  2160  &   $-$13.39   \\
RFGC~560 & 023631.6 $+$071834 & 2.89 &  1.12 &  6122 &    6723 &  2160  &   $-$12.91   \\
RFGC~603 & 025017.5 $-$083550 & 2.55 &  1.01 &  5326 &    6643 &  2160  &   $-$13.42   \\
RFGC~620 & 025426.2 $+$423900 & 2.43 &  1.21 &  2162 &    6603 &  1800  &   $-$13.11   \\
RFGC~722 & 032524.8 $-$161405 & 3.23 &  1.06 &  1873 &    6603 &  2400  &   $-$12.81   \\
RFGC~798 & 040048.9 $+$350049 & 2.55 &  1.06 &  4157 &    6643 &  2160  &   $-$12.93    \\
RFGC~855 & 042921.8 $-$044535 & 2.12 &  1.03 &  4353 &    6643 &  3600  &   $-$13.03    \\
RFGC~911 & 045146.0 $+$034005 & 2.02 &  1.00 &  4578 &    6643 &  2160  &   $-$13.41    \\
RFGC~944 & 050732.0 $-$113905 & 2.26 &  1.12 &  2358 &    6603 &  2400  &   $-$13.39    \\
RFGC~1133& 064854.0 $+$661540 & 2.24 &  1.01 &  3304 &    6643 &  3600  &   $-$12.76    \\
RFGC~1339& 081357.6 $+$523853 & 4.87 &  1.09 &  5459 &    6683 &  3600  &   $-$13.31    \\
RFGC~1434& 084850.8 $+$295212 & 2.13 &  1.05 &  5964 &    6683 &  3600  &   $-$13.57    \\
RFGC~1462& 085901.0 $+$391233 & 4.14 &  1.00 &   595 &    6603 &  2400  &   $-$12.98    \\
RFGC~1504& 091154.6 $-$200700 & 4.76 &  1.19 &  2177 &    6603 &  2400  &   $-$12.95    \\
RFGC~1700& 100236.0 $-$060049 & 4.31 &  1.16 &   661 &    6603 &  3600  &   $-$12.90    \\
RFGC~3359& 182402.4 $+$651822 & 2.52 &  1.22 &  7124 &    6723 &  2160  &   $-$13.20    \\
RFGC~3378& 183339.5 $+$320822 & 1.95 &  1.25 &  5456 &    6683 &  2400  &   $-$13.62    \\
RFGC~3385& 183754.4 $+$173201 & 2.63 &  1.14 &  4500 &    6683 &  1800  &   $-$13.05    \\
RFGC~3608& 203523.7 $-$061440 & 2.11 &  1.07 &  5798 &    6643 &  2400  &   $-$13.39    \\
RFGC~3645& 204838.4 $-$171430 & 2.08 &  1.32 &  8336 &    6723 &  2160  &   $-$13.50    \\
RFGC~3651& 204952.2 $-$070119 & 3.47 &  1.05 &  6047 &    6723 &  2160  &   $-$12.94    \\
RFGC~3803& 214439.4 $-$064121 & 2.06 &  1.27 &  3090 &    6643 &  2400  &   $-$13.39    \\
RFGC~3824& 215235.8 $+$281823 & 2.08 &  1.09 &  3476 &    6643 &  2160  &   $-$12.88    \\
RFGC~3827& 215245.5 $+$385611 & 3.09 &  1.11 &  5989 &    6723 &  2160  &   $-$12.95    \\
RFGC~3846& 215807.4 $+$010032 & 3.47 &  1.13 &  3011 &    6643 &  2160  &   $-$13.17    \\
RFGC~3880 &220804.8 $-$101959  &2.16  & 1.33  & 2866  &   6643  & 2400   &  $-$13.63    \\
RFGC~3935 &222316.6 $-$285851  &3.64  & 1.03  & 1808  &   6603  & 2700   &  $-$12.68    \\
RFGC~4039 &225912.8 $+$133624  &3.44  & 1.24  & 2568  &   6643  & 2160   &  $-$12.84    \\
RFGC~4072 &230754.9 $+$050940  &1.90  & 1.02  & 3523  &   6643  & 2160   &  $-$13.26    \\
RFGC~4078 &231203.6 $+$484859  &1.93  & 1.29  & 8657  &   6723  & 2160   &  $-$13.25    \\
RFGC~4081 &231313.1 $+$062548  &4.70  & 1.02  & 4839  &   6683  & 1800   &  $-$13.23    \\
RFGC~4091 &231502.6 $+$012608  &2.11  & 1.05  & 4961  &   6643  & 2160   &  $-$13.41    \\
RFGC~4106 &231930.4 $+$160429  &3.25  & 1.06  & 7238  &   6723  & 2160   &  $-$12.97    \\
RFGC~4149& 233543.6 $+$322306 & 2.37 &  1.12 &  4957 &    6683 &  2100  &   $-$12.95    \\ \hline
\end{tabular}
\end{table*}
 \renewcommand{\baselinestretch}{1.0}

%% file: Kaisin_tab2.tex
\renewcommand{\baselinestretch}{0.7}
\begin{turnpage}
\begin{table*}
\caption{General parameters of the UF galaxies} \label{table2}
\medskip
\begin{tabular}{r|c|c|c|c|r|r|c|c|c|r|c|r|r|r|r|c} \hline
 RFGC & T &$\log r_{25}$& $B_t$& $A_G$& $D$~~ &$V_m$&$m_{21}$ &$\log F_{{\rm H}\alpha}$ &$m_{FUV}$ &$\log M_{\rm HI}$ &$A_B$ & $m_K$~ &$\log L_K$ &$\log SFR\alpha$ &$\log SFRu$ &$\log sSFR\alpha$\\
 \hline
(1)~~~~ &(2)& (3)& (4)& (5) & (6)~ & (7) & (8)& (9)& (10)& (11)~~~~& (12) & (13)~~& (14)~~~~& (15)~~~~~~& (16)~~~~~~& (17) \\
\hline
    1 & 5 & 0.85 & 15.65 & 0.34  & 97 & 204 &  15.53  & $-$13.45 & 19.18  & 10.09 & 1.29 & 10.67 & 11.01 &  0.14  & 0.34 & $-$10.87 \\
   99 & 5 & 0.88 & 15.69 & 0.11  & 75 & 190 &  15.02  & $-$12.97 & 17.26  & 10.07 & 1.28 & 10.95 & 10.68 &  0.34  & 0.70 & $-$10.34  \\
  124 & 5 & 0.82 & 14.61 & 0.09  &101 & 294 &  16.54  & $-$12.72 & 19.92  &  9.72 & 1.51 &  9.66 & 11.45 &  0.94  & 0.06 & $-$10.51 \\
  161 & 7 & 0.82 & 14.33 & 0.09  & 22 &  86 &  14.33  & $-$12.80 &        &  9.28 & 0.63 & 10.76 &  9.69 & $-$0.84  &      & $-$10.53 \\
  176 & 6 & 0.89 & 14.77 & 0.29  & 75 & 165 &  14.93  & $-$12.98 & 17.67  & 10.11 & 1.19 & 10.19 & 10.98 &  0.33  & 0.60 & $-$10.65 \\
  255 & 7 & 0.73 & 14.67 & 0.11  & 24 &  89 &  14.48  & $-$12.77 & 16.41  &  9.30 & 0.58 & 11.13 &  9.62 & $-$0.75  &$-$0.49 & $-$10.37 \\
  438 & 7 & 1.04 & 14.61 & 0.11  & 53 & 117 &  14.88  & $-$12.73 & 16.71  &  9.83 & 1.08 & 10.57 & 10.53 &  0.19  & 0.46 & $-$10.34 \\
  463 & 6 & 0.73 & 16.08 & 0.21  & 66 & 104 &  15.77  & $-$13.46 &        &  9.66 & 0.68 & 12.09 & 10.11 & $-$0.50  &      & $-$10.61  \\
  504 & 6 & 0.86 & 14.73 & 0.66  & 53 & 180 &  15.03  & $-$12.85 & 18.28  &  9.77 & 1.21 &  9.76 & 10.85 &  0.26  & 0.36 & $-$10.59   \\
  511 & 5 & 1.00 & 15.88 & 0.13  &131 &     &         & $-$13.61 & 18.90  &       & 1.50 & 10.98 & 11.12 &  0.28  & 0.71 & $-$10.84  \\
  517 & 5 & 0.86 & 15.41 & 0.40  & 74 & 128 &  16.15  & $-$13.21 & 19.66  &  9.61 & 1.07 & 10.59 & 10.81 &  0.07  &$-$0.21 & $-$10.74  \\
  531 & 6 & 0.75 & 16.0  & 1.00  & 58 &  93 &  15.67  & $-$13.39 & 20.06  &  9.59 & 0.63 & 11.27 & 10.33 & $-$0.37  &$-$0.46 & $-$10.70  \\
  560 & 5 & 0.90 & 15.44 & 0.54  & 85 & 202 &  14.74  & $-$12.91 & 18.71  & 10.29 & 1.36 & 10.19 & 11.09 &  0.64  & 0.62 & $-$10.45  \\
  603 & 5 & 0.95 & 15.02 & 0.13  & 73 & 130 &  16.08  & $-$13.42 & 18.52  &  9.62 & 1.07 & 10.47 & 10.85 & $-$0.22  & 0.02 & $-$11.07  \\
  620 & 7 & 0.69 & 15.63 & 0.40  & 32 &  86 &  15.53  & $-$13.11 & 18.47  &  9.13 & 0.53 & 11.85 &  9.58 & $-$0.79  &$-$0.88 & $-$10.37  \\
  722 & 7 & 0.83 & 15.33 & 0.18  & 25 & 101 &  14.12  & $-$12.81 & 16.61  &  9.48 & 0.75 & 11.55 &  9.48 & $-$0.67  &$-$0.35 & $-$10.15   \\
  798 & 5 & 0.89 & 15.19 & 0.92  & 59 & 166 &  15.20  & $-$12.93 &        &  9.79 & 1.19 &  9.73 & 10.96 &  0.33  &      & $-$10.63  \\
  855 & 6 & 0.92 & 15.38 & 0.23  & 59 & 136 &  15.40  & $-$13.03 & 18.09  &  9.71 & 1.07 & 10.98 & 10.46 &  0.01  & 0.09 & $-$10.45  \\
  911 & 6 & 0.94 & 16.0  & 0.27  & 62 & 111 &  14.68  & $-$13.41 &        & 10.04 & 0.93 & 11.70 & 10.21 & $-$0.38  &      & $-$10.59  \\
  944 & 7 & 0.75 & 15.40 & 0.69  & 31 &  99 &  15.48  & $-$13.39 &        &  9.12 & 0.67 & 11.19 &  9.82 & $-$0.97  &      & $-$10.79  \\
 1133 & 4 & 0.98 & 15.28 & 0.35  & 47 & 142 &  15.34  & $-$12.76 &        &  9.54 & 1.18 & 10.15 & 10.59 &  0.16  &      & $-$10.43  \\
 1339 & 5 & 0.89 & 15.23 & 0.19  & 74 & 277 &  14.52  & $-$13.31 & 19.27  & 10.26 & 1.59 & 10.10 & 11.01 &  0.14  & 0.18 & $-$10.87  \\
 1434 & 5 & 0.69 & 15.52 & 0.17  & 81 & 186 &  15.66  & $-$13.57 & 19.00  &  9.88 & 0.99 & 11.01 & 10.72 & $-$0.31  &$-$0.11 & $-$11.03   \\
 1462 & 7 & 0.98 & 15.18 & 0.13  & 15 &  48 &  14.39  & $-$12.98 & 16.21  &  8.93 & 0.26 & 11.94 &  8.89 & $-$1.50  &$-$1.05 & $-$10.39  \\
 1504 & 5 & 1.14 & 14.74 & 0.66  & 26 & 156 &  13.82  & $-$12.95 & 18.18  &  9.63 & 1.47 &  9.26 & 10.44 & $-$0.35  &$-$0.02 & $-$10.79  \\
 1700 & 7 & 1.15 & 14.70 & 0.17  & 10 &  53 &  14.40  & $-$12.90 & 16.60  &  8.57 & 0.40 & 11.28 &  8.80 & $-$1.71  &$-$1.42 & $-$10.51  \\
 3359 & 5 & 1.01 & 15.76 & 0.18  &101 & 191 &  15.40  & $-$13.20 & 18.96  & 10.18 & 1.48 & 10.75 & 11.02 &  0.47  & 0.49 & $-$10.55  \\
 3378 & 5 & 1.15 & 16.50 & 0.38  & 78 & 184 &  16.11  & $-$13.62 & 19.82  &  9.67 & 1.64 & 11.13 & 10.64 & $-$0.06  & 0.19 & $-$10.70  \\
 \hline
 \end{tabular}
\end{table*}
\end{turnpage}

\begin{turnpage}
\setcounter{table}{1}
\begin{table*}[]
 \caption{Contd.}
\medskip
\begin{tabular}{r|c|c|c|c|r|r|c|c|c|r|c|r|r|r|r|c} \hline
 RFGC & T &$\log r_{25}$& $B_t$& $A_G$& $D$~~ &$V_m$&$m_{21}$ &$\log F_{{\rm H}\alpha}$ &$m_{FUV}$ &$\log M_{\rm HI}$ &$A_B$ & $m_K$~ &$\log L_K$ &$\log SFR\alpha$ &$\log SFRu$ &$\log sSFR\alpha$\\
 \hline
(1)~~~~ &(2)& (3)& (4)& (5) & (6)~ & (7) & (8)& (9)& (10)& (11)~~~~& (12) & (13)~~& (14)~~~~& (15)~~~~~~& (16)~~~~~~& (17) \\
\hline
 3385 & 5 & 0.74 & 15.39 & 1.27  & 65 & 234 &  14.75  & $-$13.05 &        & 10.06 & 1.21 &  9.56 & 11.11 &  0.38  &      & $-$10.73  \\
 3608 & 5 & 0.96 & 15.38 & 0.22  & 82 & 278 &  15.44  & $-$12.92 & 19.06  &  9.98 & 1.72 & 10.09 & 11.10 &  0.68  & 0.48 & $-$10.42   \\
 3645 & 6 & 1.03 & 16.1  & 0.25  &116 &     &         & $-$13.50 &        &       & 1.40 & 11.39 & 10.86 &  0.27  &      & $-$10.59  \\
 3651 & 5 & 0.90 & 15.15 & 0.27  & 85 & 253 &  14.80  & $-$12.94 & 18.03  & 10.27 & 1.54 &  9.99 & 11.17 &  0.62  & 0.82 & $-$10.55  \\
 3803 & 6 & 0.99 & 15.70 & 0.16  & 45 &  89 &  16.11  & $-$13.39 & 17.58  &  9.19 & 0.79 & 11.65 &  9.96 & $-$0.72  &$-$0.21 & $-$10.68  \\
 3824 & 6 & 0.94 & 15.28 & 0.36  & 52 & 113 &  15.22  & $-$12.88 &        &  9.67 & 0.95 & 10.87 & 10.39 &  0.03  &      & $-$10.36  \\
 3827 & 5 & 1.12 & 16.50 & 1.39  & 86 & 231 &  15.55  & $-$13.25 &        &  9.98 & 1.82 &  9.94 & 11.20 &  0.72  &      & $-$10.48  \\
 3846 & 4 & 0.79 & 15.48 & 0.23  & 44 & 143 &  14.69  & $-$13.17 & 18.41  &  9.74 & 0.96 & 10.69 & 10.32 & $-$0.43  &$-$0.48 & $-$10.75 \\
 3880 & 7 & 1.09 & 16.0  & 0.16  & 42 &  84 &  15.72  & $-$13.63 & 18.03  &  9.29 & 0.82 & 12.17 &  9.69 & $-$1.01  &$-$0.43 & $-$10.70  \\
 3935 & 7 & 0.93 & 14.36 & 0.08  & 19 &  60 &  14.24  & $-$12.68 & 16.10  &  9.19 & 0.42 & 11.01 &  9.46 & $-$0.94  &$-$0.72 & $-$10.40 \\
 4039 & 6 & 0.82 & 14.92 & 0.35  & 38 & 121 &  14.23  & $-$12.84 &        &  9.80 & 0.87 & 10.60 & 10.23 & $-$0.24  &      & $-$10.47 \\
 4072 & 5 & 0.67 & 16.39 & 0.29  & 51 & 113 &  15.63  & $-$13.26 & 19.03  &  9.49 & 0.67 & 12.08 &  9.89 & $-$0.51  &$-$0.68 & $-$10.40 \\
 4078 & 5 & 1.22 & 16.50 & 1.00  &123 & 269 &  15.83  & $-$13.55 &        & 10.18 & 2.14 & 10.01 & 11.48 &  0.77  &      & $-$10.71 \\
 4081 & 5 & 0.95 & 14.53 & 0.47  & 70 & 236 &  14.00  & $-$13.23 & 17.90  & 10.42 & 1.56 &  9.15 & 11.34 &  0.22  & 0.88 & $-$11.12 \\
 4091 & 5 & 0.94 & 15.57 & 0.24  & 71 & 137 &  15.36  & $-$13.41 & 18.51  &  9.89 & 1.10 & 10.88 & 10.66 & $-$0.20  & 0.11 & $-$10.86 \\
 4106 & 5 & 0.67 & 15.51 & 0.24  &102 & 217 &  15.12  & $-$12.97 & 18.04  & 10.30 & 1.05 & 10.87 & 10.98 &  0.54  & 0.58 & $-$10.44  \\
 4149 & 6 & 0.70 & 15.54 & 0.35  & 72 & 100 &  14.78  & $-$12.95 & 17.31  & 10.13 & 0.63 & 11.46 & 10.44 &  0.10  & 0.33 & $-$10.34 \\
\hline
 2246 & 7 & 1.22 & 14.10 & 0.12  & 17 &  98 &  13.70  & $-$12.69 & 16.30  &  9.31 & 1.08 & 10.05 &  9.75 & $-$0.76  &$-$0.35 & $-$10.51 \\
  626 & 7 & 0.85 & 16.93 & 0.60  & 91 & 101 &  16.23  & $-$13.77 &        &  9.76 & 0.77 & 12.71 & 10.14 & $-$0.40  &      & $-$10.54 \\
 1446 & 5 & 0.85 & 16.40 & 0.19  & 82 & 126 &  16.32  & $-$13.36 & 19.46  &  9.63 & 0.94 & 11.92 & 10.37 & $-$0.10  &$-$0.30 & $-$10.47 \\
251824& 5 & 1.00 & 17.52 & 0.17  & 88 &  93 &  16.65  & $-$13.80 & 19.43  &  9.56 & 0.84 & 13.16 &  9.93 & $-$0.53  &$-$0.32 & $-$10.46  \\
 2026 & 5 & 0.90 & 15.75 & 0.08  &104 & 218 &  16.10  & $-$13.34 & 19.85  &  9.92 & 1.42 & 10.90 & 10.98 &  0.30  & 0.03 & $-$10.68   \\
 2079 & 5 & 0.80 & 16.71 & 0.10  & 94 & 158 &  16.30  & $-$13.34 & 19.77  &  9.76 & 1.04 & 12.22 & 10.37 &  0.06  &$-$0.30 & $-$10.31  \\
 2253 & 5 & 0.99 & 16.65 & 0.12  & 91 & 160 &  16.98  & $-$13.43 & 20.73  &  9.46 & 1.30 & 11.88 & 10.48 &  0.06  &$-$0.50 & $-$10.42  \\
 2322 & 5 & 0.71 & 16.07 & 0.07  & 94 & 164 &  15.93  & $-$13.33 & 18.31  &  9.90 & 0.94 & 11.71 & 10.57 &  0.02  & 0.18 & $-$10.55  \\
 2339 & 5 & 0.82 & 15.87 & 0.09  &100 & 180 &  15.28  & $-$13.38 & 18.57  & 10.22 & 1.16 & 11.27 & 10.80 &  0.12  & 0.32 & $-$10.68  \\
 2461 & 6 & 0.80 & 17.00 & 0.07  & 97 & 116 &  16.62  & $-$13.51 & 19.63  &  9.66 & 0.82 & 13.01 & 10.08 & $-$0.19  &$-$0.41 & $-$10.27  \\
\hline
\end{tabular}
\end{table*}
\end{turnpage}
\renewcommand{\baselinestretch}{1.0}